Title : **Posterior Basicranium asymmetry and idiopathic scoliosis**

**Authors** :D. L. Rousie, Docteur en Neurosciences (1), O. Joly, PhD student (1), A. Berthoz, Professeur au College de France (1), ( (1) Laboratoire de la perception et de l action, College de France, Paris)

## Comments

Name and adress for correspondence : Docteur Rousie, 3 rue Saint Louis, 59113 Seclin France. Fax: 0033 320 32 35 44, Tel. 0033 320 90 12 29, mrousie@nordnet.fr

## Supports :

## Key points :

Posterior Basicranium (PBA*),* divided in 2 parts joined to the foramen magnum, presents a torque in Idiopathic scoliosis (IS)
PBA torque reflects cerebellar asymmetry in IS.
PBA torque induces spatial asymmetry of the vestibular system and specially, of the otolithic system, with potential consequences on vestibular function and posture in IS.

## Abstract :

PBA asymmetry and idiopathic scoliosis
Study design
Are there neuro-anatomical abnormalities associated with idiopathic scoliosis (IS)?
Posterior Basicranium (PBA) reflects cerebellum growth and contains vestibular organs, two structures suspected to be involved in scoliosis.
Objective
The aim of this study was to compare posterior basicranium asymmetry (PBA) in Idiopathic scoliosis (IS) and normal subjects.
Method: To measure the shape of PBA in 3D, we defined an intra-cranial frame of reference based on CNS and guided by embryology of the neural tube. Measurements concerned three directions of space referred to a specific intra cranial referential. Data acquisition was performed with T2 MRI (G.E. Excite 1.5T, mode Fiesta). We explored a scoliosis group of 76 women and 20 men with a mean age of 17, 2 and a control group of 26 women and 16 men, with a mean age of 27, 7.
Results: IS revealed a significant asymmetry of PBA ($Pr>|t|<.0001$) in 3 directions of space compared to the control group. This asymmetry was more pronounced in antero-posterior (AP) & lateral direction, forming a torque of the posterior base shape associated with identical cerebellar torque.
Discussion: IS Cerebellar and PBA asymmetries are neurodevelopmental anomalies involved in IS. The vestibular systems are embedded in the 2-parted PBA and, therefore, are in asymmetrical spatial position with potential consequences on otolithic function because this latter is devoted to gravity.
In 2002, Kwak demonstrated links between patterning of the cerebellum and alterations in the development of the internal ear in experimental zebra fish (focused on semi-circular canals formation) We engaged a study on IS semi-circular canals anatomy thanks to a novel program of modelling. We will report our first results in the companion paper.

# Pba asymmetry and idiopathic scoliosis

## Introduction

According to a report by Nachemson[1],idiopathic Scoliosis (IS) is now generally recognized to be caused by several conditions: genetic, [2, 3], hormonal [4, 5], neuro-anatomical [6, 7, 8] and/or neuro-muscular [9, 10].
Various studies have reported links between IS and asymmetry at similar functional and anatomical levels: Chockalingam [11], Shimode[12], Kramers[13]and Perret[10] have reported right/left asymmetries at functional levels for gait or motor control parameters.
Geissele [14] and Kertesz[15] have stressed the presence of links between IS and right/left brain asymmetries. Geschwind & Galaburda [16] have highlighted asymmetries of the temporal planum, with the left planum being more developed in 65% of examined brains.
In a study by Lundstrom [17] 60 to75% of the examinated population has a left hemisphere larger than the right. He measured the angle formed by a line drawn from the ear to the median line. This anatomical detail has also been noted by Burke [18] who pointed out that cause of this asymmetry is still unknown. More recently, neuro-anatomical abnormalities in the pontine and hindbrain regions have been clearly implicated, in IS, in a report by Lowe [19] in 2000.
Pirttiniemi, Lahtela, Hunggare & Serlo[20], Delaire[21], Berthoz &Rousié [22,23] have reported the presence of craniofacial asymmetry(CFA) in certain cases of scoliosis and /or laterocolis. These asymmetries are associated with spatial asymmetries of vestibular organs involved in postural control.
CFA implicates asymmetry of each bony component of the head: vault, face and basicranium. The basicranium is divided in three parts: the anterior part, the middle part which is a true patella which makes the 3D movements of the two other parts possible and the posterior part in which the labyrinths are embedded.
The basicranium shape reflects the underlying brain growth. Cerebral asymmetry appears, in utero,at the beginning of the third week. At this stage, the basicranium is a pre-cartilaginous and flexible structure: the three parts of the base are separated by large and mesenchymal synchondrosis avoiding any pressure that could affect the brain growth [24, 25]. It is therefore obvious that no basicranium asymmetry would occur without cerebral asymmetry.
The aim of this study was to compare the shape of the posterior basicranium (PBA) on IS and non-IS subjects. This study was conducted for two reasons: 1) the PBA reflects the growth of the cerebellum, the asymmetry of which is though to be implicated in scoliosis [19, 26] and 2) the PBA contains vestibular organs implicated in postural disorders [27].

## Materials and Methods

MRI acquisition and measurement of basicranium asymmetries
Data, in all MRI machines are included in a volume, which is itself included in the reference frame of the machine. Three dimensional-coordinates of a selected point (voxel) are calculated from this frame and are <u>dependent on the patient's head orientation in the machine.</u>

Precise measurements require a reliable frame of reference, apart from the patient head orientation during the acquisition. These measurements should allow comparison of the spatial location of head paired structures in a subject but also between different subjects.

We used an intra-cranial frame of reference based on the central nervous system [28]. We assigned our reference points on the basis of embryology: the neural tube stage marks the beginning of the cranio-caudal orientation of the embryo especially at the cerebral level. This stage is also the epicenter of brain-lobes growth and at this stage, asymmetry is not yet identify.

The neuro sagittal medial plan (NSM), (red plan in Figure 1)-is built from points located on the axis upon which this embryonic neural tube was fastened: two points are taken successively along on the fundus of the third-ventricle (mesencephalon) and one point is taken at the level of the cerebral aqueduct joining the third to the fourth ventricle (see Figure 2A, B, C, D)

The axial plan-(blue plan in Figure 1)- is perpendicular to NSM plan and goes through the CA (anterior commissure)-to CP (posterior commissure) axis which provides the principal point of reference in Talairach's atlas [29]. Thus, to build this plan, we used the three following points: CA, CP and the point where CA-CP axis crosses the sagittal plan.

The frontal plan-(green plan in Figure 1B) - is perpendicular to the previous ones.

After constructing this frame of reference; it became possible to compare the 3D location of paired structures of the head.

The PBA is divided into two parts: the left and right petrous pyramids joined to the foramen magnum. We used the junction between the auditory meatus and the three vestibular semi-circular canals as markers to measure the spatial position of each part. These points, named P (right) & P' (left)-and situated inside the PBA, are precise and easy to find. (See Figure 3A, B)

Data acquisition was performed with EXCITE MRI 1.5T., from General Electric, using a head coil. A T2-weighted sequence was used in 3D FIESTA mode acquisition Thus, PBA structures were analysed in a volume instead of classical MRI slices. The quality of data was monitored by a G.E. work station 4.1 and stored on a CD.

We processed the data using various modules of Brainvisa which is free access software devoted to MRI available on the web (http://brainvisa.info/).

Various steps of the process:

1-Import and convert MRI data (Dicom format) given by the GE machine to 'Gis Brain visa' format.

2- Select and save points needed for the construction of the reference frame. For each chosen point, a module of Brainvisa called Anatomist 1.3 provides 3D coordinates referenced to the reference frame of the MRI machine.

3- Select and save the markers points P (right) and P' (left).

4- Plans and distances :we automated the following procedure in a novel program in order to measure the 3D location of P&P 'referenced on our reference frame: in a mathematical method, we calculate the equation of the planes using Cartesian coordinates of three distinct points : $A(x_A, y_A, z_A)$, $B(x_B, y_B, z_B)$ et $C(x_C, y_C, z_C)$

The equation of plan is: $ax + bx + cz + d = 0$. The program automatically calculates the distances of these three points to the plans.

By measuring the distance differences between P and P'-to each of the 3 planes, we can define the spatial orientation of both parts of the PBA:

1) to the NSM plane = dP_sag – dP'_sag, 2)to axial plane =dP_axial – dP'_axial, 3) to the frontal plane = dP_front – dP'front. These distances are given in millimeters.

We used a similar procedure for each patient for establishing inter-patients comparisons.

# Pba asymmetry and idiopathic scoliosis

## Patients and control subjects

No data was available in relation to PBA evaluation in IS patient at the time of the planning: thus the sample size could not be estimated before hand. The number of IS patients was fixed between 80 and 100 because some abnormalities may be uncommon in the studied IS population and also because classical statistical analyses related to categorical variables require such sample size. The number of controls was kept to 30 as the Student's t-test has been shown to be robust when the sample size is not too small.

a) Control group (CG)

Justifying a MRI for healthy individuals was difficult: thus, control subjects were recruited from patients undergoing maxillo-facial consultation and suffering from lesions of the temporomandibular joints which are similar to our regions of interest. These lesions do not interfere with spine or PBA and these patients can be included as normal control subjects. We exploited, their MRI results, which have been carried through in the context of their therapeutic treatment. The criteria for inclusion were strictly observed: patients were required to be free of any spine deformation, whether congenital or acquired which was confirmed by the rheumatologists of our team. All of these subjects were consenting to give their MRI and spine data (spine radiographies) in an anonymous way for the study.

This control group (CG) included 33 persons, 27 women and16 men, ranging in age from 8 to 51 years with a mean age of 27, 7 years (std. dev: 5.6).

106b) Scoliosis group (SG)

For the scoliotic group (SC), the exclusion criteria were to be free of acquired or congenital spine lesion, of visual or auditive malformation which could have postural incidences and of systemic illness. IS patients have been recruited from patients treated by physiotherapy and/or brace by rheumatologists of our team (hospital or clinics). The recruitment was time-randomized (23months): patients, presenting the inclusion criteria, were asked to participate to the study as they came along to consultations. MRI was prescribed, in the course of their therapeutic treatment to control malformations or lesions in the skull base or cervical spine (which could also be exclusion criteria).They gave their consent to give their MRI and spine data, in anonymous way, for the study.

SG included 95 patients, 75 females and 20 males, ranging in age from 10 to 30 with a mean age of 17.3 (Std. dev.: 4.7).The greater number of females reflects the well-known large incidence of idiopathic scoliosis on female.

Their scoliosis was classified following two criteria

1 –Cobb angle: two sub-groups were distinguished according to the severity of the curve: SG1: from 12 to 20 degrees, including 57 subjects with a mean age of 14, 8(std. dev.: 3.01). SG2: from 20 to 47degrees including 38 subjects with a mean age of 18, 9(std.dev. 3, 85).

2- Location of the deformation along the spine according to the Cottalorda and Kohler classification [30]

The randomized recruitment, spread over 23 months, allowed us to divide the same 95 patients in three sub-groups:

- 57 patients with a thoracolumbar deformation formed the TL subgroup (52 with left TL and 5 with right TL)
- 24 patients with a thoracic deformation formed the T subgroup (22 with right T and 2 with left T)
- 14 patients with a lumbar deformation formed the L subgroup (13 with left L and 1 with right L)

## Results

Reliability Analyses:

We evaluated the reliability of our method by testing the reproducibility of our results with test-retest correlation (RO-intra) using the Fermanian Intra-class correlation reported by Fleiss [31]. The Fermanian scale of evaluation is as follows: RO>0.91= very good; 0.71<RO<0.91=good; 0.51<RO<0.71= moderated.

We measured the dP-dP'distances to the NSM, frontal and vertical planes.

As recommended in the method, these measurements were carried out three times. Thus, for each subject, we collected nine measurements. This evaluation was carried out using 26 subjects.

We obtained the following results:

To the NSM, RO=0.97752 with an inferior limit=0.94739, sup. Limit=0, 99048; Estimation= very good.

To the frontal plane RO=0, 97872, inferior limit=0, 95014, sup. Limit=0, 99099; Estimation=very good.

To the vertical plane RO= 0, 93930, inferior limit=0, 86152, sup. Limit= 0, 97401; Estimation= very good.

## Statistical analysis

All statistical analysis was performed by means of SAS software (SAS Institute Inc. Cary, NC 25513). P values<0.05 were considered statistically significant. Results were expressed as the mean, standard deviation and range .The assumption of equal variances was tested using the Fisher's test. The student's test was used because this assumption was not rejected.

Comparison between normal and scoliosis Basicranium asymmetries

IS patients (SG) presented significantly greater asymmetry of the PBA orientation in 3 directions of space than the control group (CG)

-Frontal plane: in SG, dP-dP' (mean value) = 5,662mm, Pr>|t|<.0001;
in CG, dP-dP' (mean value) = 2,140mm, Pr>|t|<.0001
-NSM plane: in SG, dP-dP' (mean value) = 3,573mm, Pr>|t|<.0001
in CG, dP-dP' (mean value) = 1,945mm, Pr>|t|<.0001
-Vertical plane: in SG, dP-dP' (mean value) =3,516mm, Pr>|t|<.0001
in CG, dP-dP' (mean value) =1,264mm, Pr>|t|<.0001.

We determined the right /left asymmetry of the 2-parted PBA, by establishing the following differences: dP- dP'negative, in relation to the NSM plane, meant that P was closer than P'. dP- dP' negative in relation to the horizontal plane, meant that the P was lower and dP-dP' negative, in relation to the frontal plane, meant that P was backward. Measurements were only calculated for the SG group.

We found: the right part of the PBA referenced to the frontal plane, backward in 66/95(69, 4%). The right part of the PBA referenced to the NSM plane was laterally closer than the left one in 54/95(56, 8%). And, referenced to the horizontal plane, the right part was lower than the left one in 73/95(76, 8%). Indirectly, this PBA torque suggested that underlying cerebellum was, most of the time, larger on left side.

## Discussion

The purpose of this study was to verify the association between scoliosis and PBA.

Our first result was methodological in which we proposed a novel technique for analyzing scoliosis: an intracranial frame of reference built from the NSM plane, automated in a programme for identifying the asymmetry proved to be a reliable tool due to statistical analysis. This technique may be useful for diagnosis in radiology but also for fundamental studies in development.

The relation between scoliosis and posterior basicranium asymmetries

Our second result was in relation to our suggestion that scoliosis patients, in comparison with normal subjects are afflicted with a significant asymmetry of the posterior basicranium(figure 4). Interestingly, even with non-scoliotic control subjects, this part of the base shows weak asymmetry which is consistent with other studies of brain and skull asymmetries [32, 33] Coffin [34], reported similar observations in 1986, and speculated that this asymmetry might have started during the first three months; this is because at this stage, the basicranium is still flexible and open to changes in its shape depending on brain growth. The "normal pattern" we obtained, provides the limits for normality and a preliminary scale of comparison. However, it would be necessary to increase the number of subjects to refine the pattern exhibited by our study with other similar groups.

IS versus control subjects revealed a significant antero-posterior and weak dorso-ventral movement of the posterior base following asymmetrical brain growth.

Our results were also in accord those of Villemann (1976) [35] who confirmed that impact of neural mass growth can modify the spatial orientation of the basicranium; In our study, we highlighted that the two aisles of PBA form a torque around the foramen magnum. Recent MRI scans confirm our findings: in 2006, Lancefield [36] found that brain torque may be observed in the latter part of foetal life. IS patients show an exaggeration of this torque. Similarly, Lacy and Horner (1996) have also demonstrated that "asymmetry" can be considered as a threshold phenomenon in cases of genetic and /or environmental modifications: they observed that skull and skeletal asymmetries increased with genetic transmission in the last few populations of rats that have been maintained under abnormal living conditions [37]. Morever, Lowe [19] has suggested a possible implication of the midbrain or hindbrain in IS. The cerebellar torque we found is also associated with an asymmetry between the right and left part of the cerebellum. A question remains whether the cerebellum asymmetry could be involved through its spatial organisation or its cortical function. Further studies could be engaged.

In the method we differentiated IS group in SG1 and SG2 subgroups for a better presentation of IS patients. One could wonder if the severity of the spine deformation is related to the amplitude of the PBA torque. We did not lead this comparison because, in the course of the study, we discovered a bias between these two subgroups: the mean age of the patients in SG1 and SG2 was, respectively, 14.8 and 18.9. In SG1, patients were younger and IS were still progressive forbidding the comparison. We should lead linear study to shed some light on this question.

## Pba asymmetry and idiopathic scoliosis

This study linked face asymmetry to basicranium asymmetry: temporo-mandibular joints hang the mandible to the posterior basicranium. This is expressed on the face and can easily be verify visually by simply assessing the implantation of the IS patient's ears and eyes during the clinical examination of the face. At the bones level of the face, PBA frequently,induces asymmetry in orbits location, deviation of the nasal septum, jaw and zygomatic arch asymmetry. This observation was previously reported by Burke 1992[18] Chebib and Chamma 1981[38], Lundstrom 1961[17] and especially Previc in 1991 [39] who carried out an extensive study of cranio-facial asymmetries confirming the association between scoliosis and cephalic asymmetry – although without making use of any other comparative methods.

The second part of our study on the PBA concentrated on the vestibular system which is widely though to be implicated in IS. The asymmetrical spatialization of the PBA aisles in which the labyrinths are embedded, suggests functional consequences as a result on vestibular function. The otolithic system which is a gravity devoted system could be responsible for passing asymmetrical information to the postural system. Previc (1991) (40),Bacsi (2004)(27) have stressed the contribution of vestibular asymmetry in postural disorders and on the asymmetry of the vestibulospinal reflex. Recently, Burwell (2006) [8] suggested that neurodevelopmental anomalies were implicated in CNS body schema and may be the primary in of idiopathic scoliosis. We are currently studying IS otolithic function

IS cerebellum and PBA asymmetries are neurodevelopmental anomalies. In 2002, Kwak[40] showed with zebra fish that the experimental 'Valentino'mutation disturbed the patterning of the cerebellum with, as a secondary consequence, alterations in the development of the internal ear due to poor expression of the fibroblastic factor fgf3. In 2004, Philips [41] confirmed the implication of factors fgf3 and fgf8 in the formation of the cerebellum and otic placodes. More, Riley in 2003, has discovered that cells abutting the posterior lateral hindbrain coordinates the early patterning of the inner ear giving to the fore the close relation between the hindbrain and the vestibular system [42]

The fact that fibroblastic factors figure equally in the formation of inner ear, cerebellum and Base suggests an identical co-involvement, in IS. We therefore designed a novel method for visualizing the anatomy of the vestibular system and more precisely that of the membranous semi-circular canals. The first results will be published in a compagnon paper.

## Titles & legends of figures

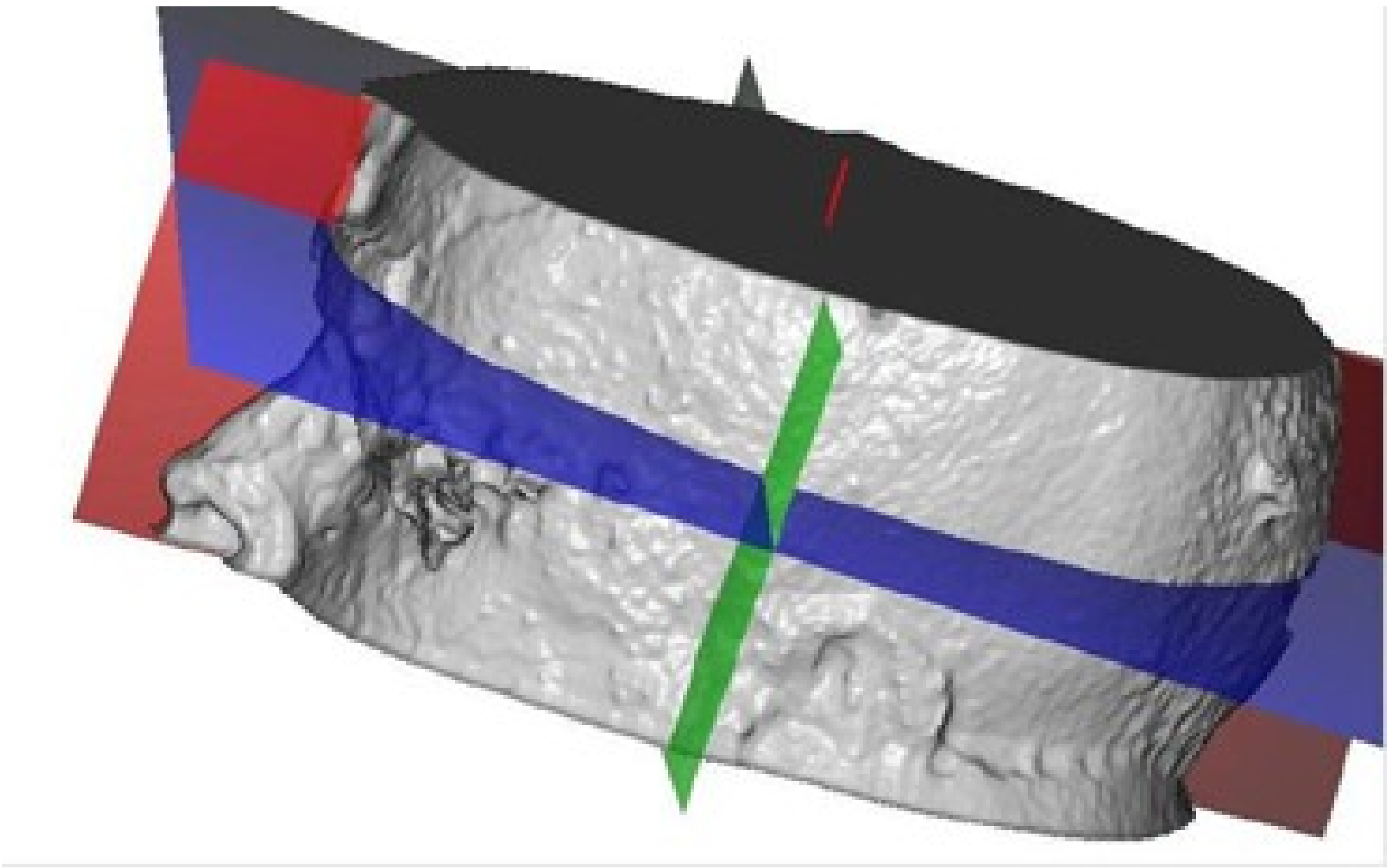

Figure 1 the intracranial frame of reference

View of the referential included in a 3D reconstruction of the head. With this referential the measurements are independent of the head's patient orientation during MRI and it allows inter-subjects PBA comparisons.

Red plane=sagittal plane used for lateral measurements, blue plane= axial plane used for vertical measurements, green plane= frontal plane used for anteroposterior measurements

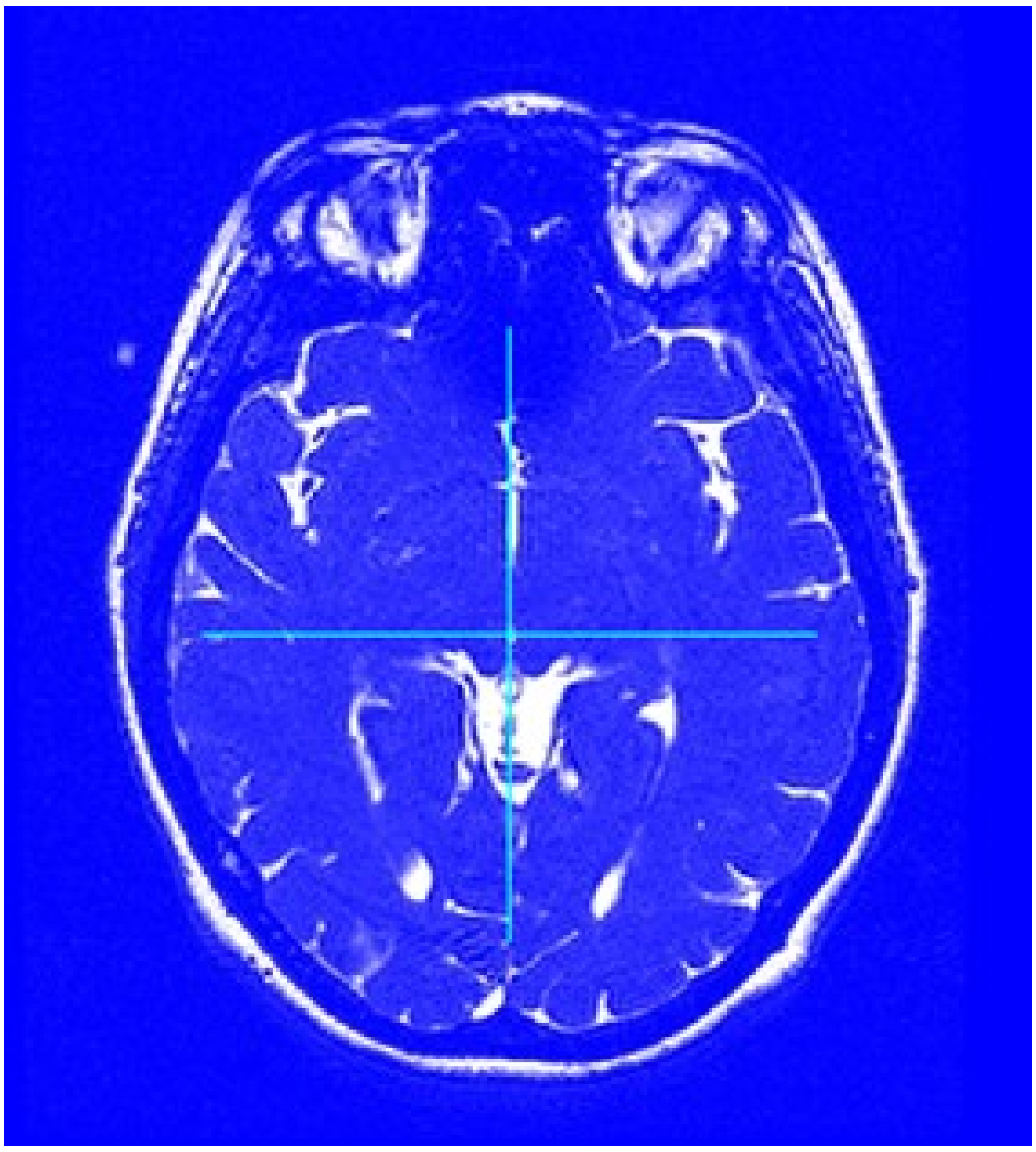

Figure 2 : A Axial view
The fundus of the third ventricle is easily identified on this axial view. The crossing lines point corresponds to the first selected point located on the third ventricle. Its 3D coordinates are automatically transmitted to the programme.

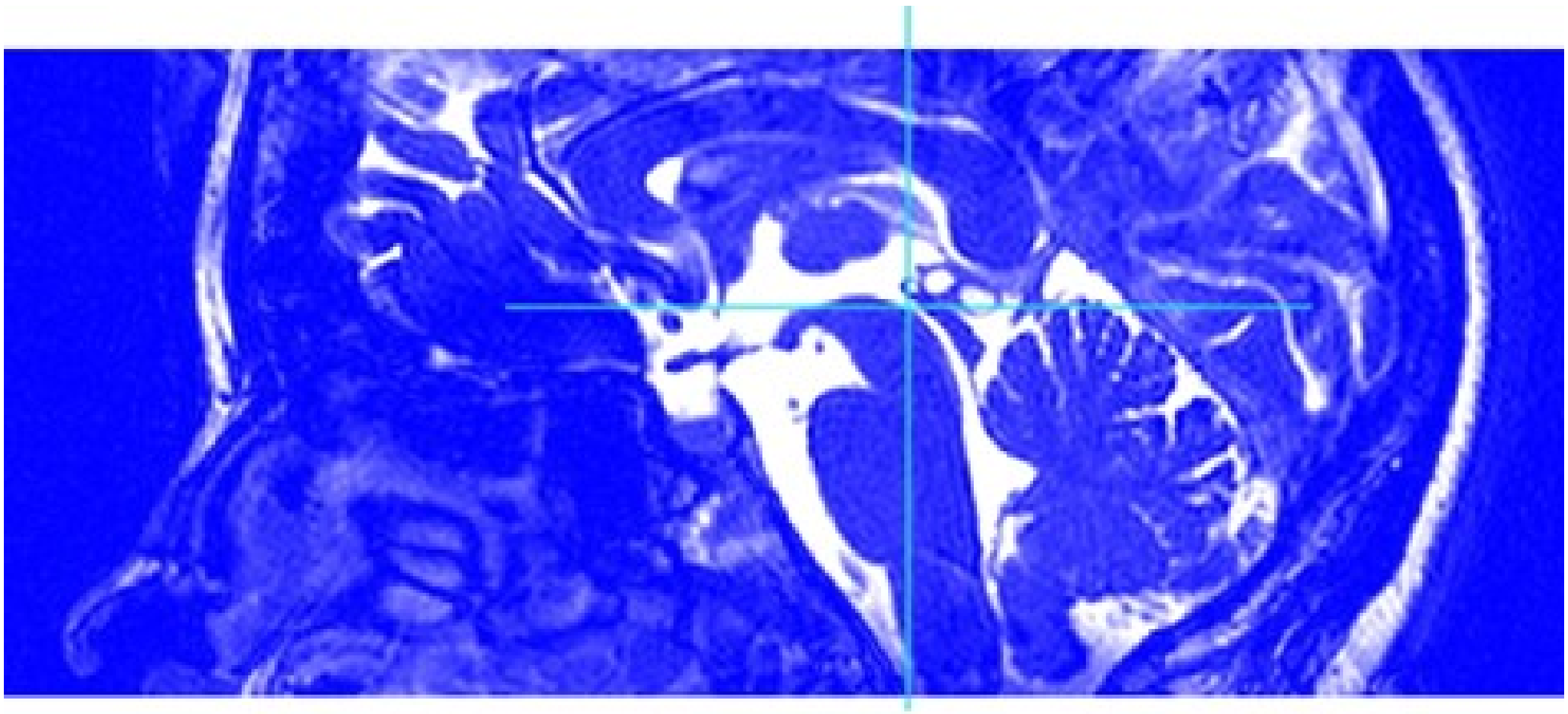

Figure 2B : Medial-sagittal view
On this medial-sagittal view the same selected point appears on the fundus of the third ventricle. This view permits to control the validity of our selected point.

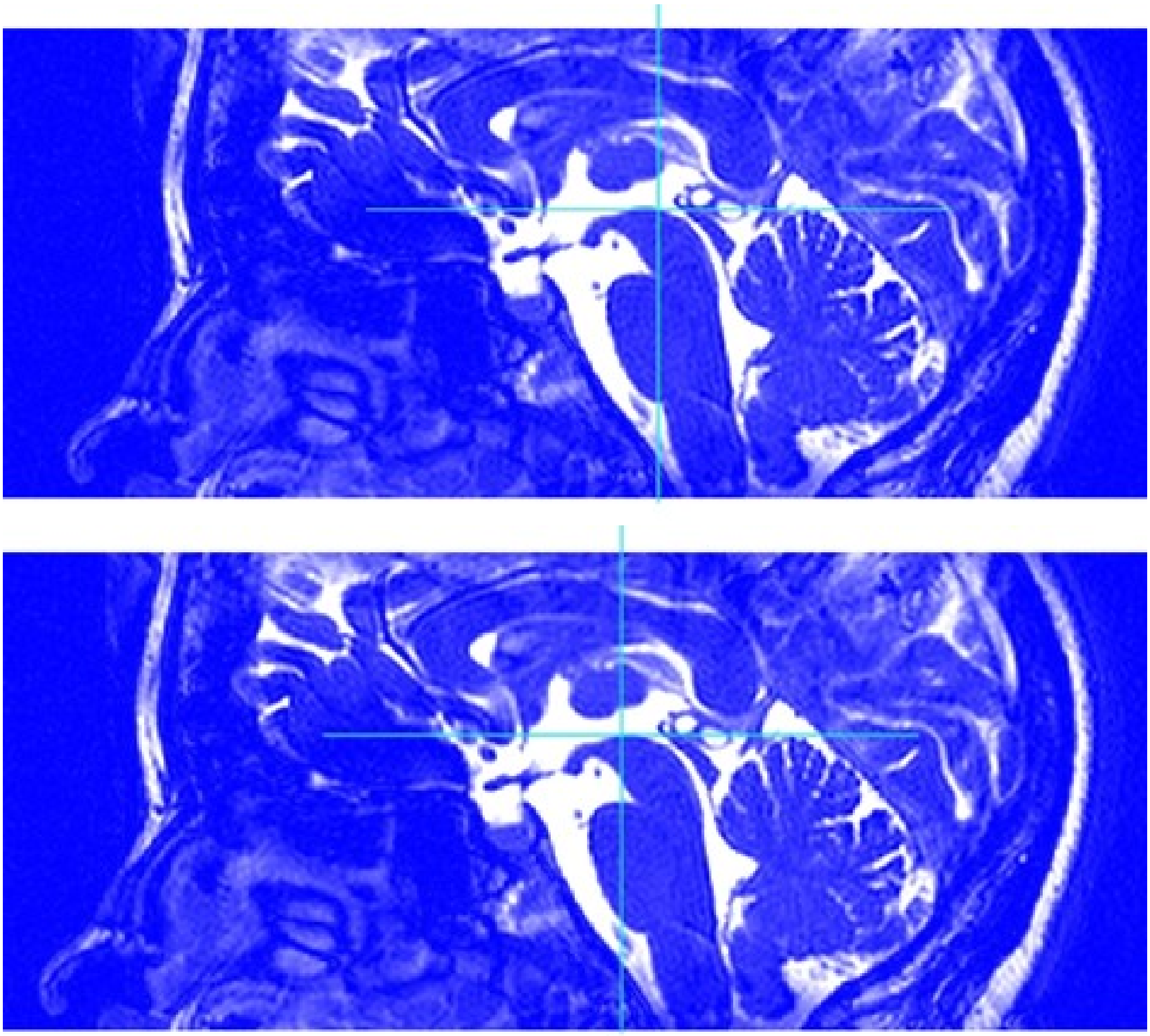

Figure 2C & 2D : Medial-sagittal view and selected points
These two views show the two others selected points which are required to create the medial axial plane of the referential.

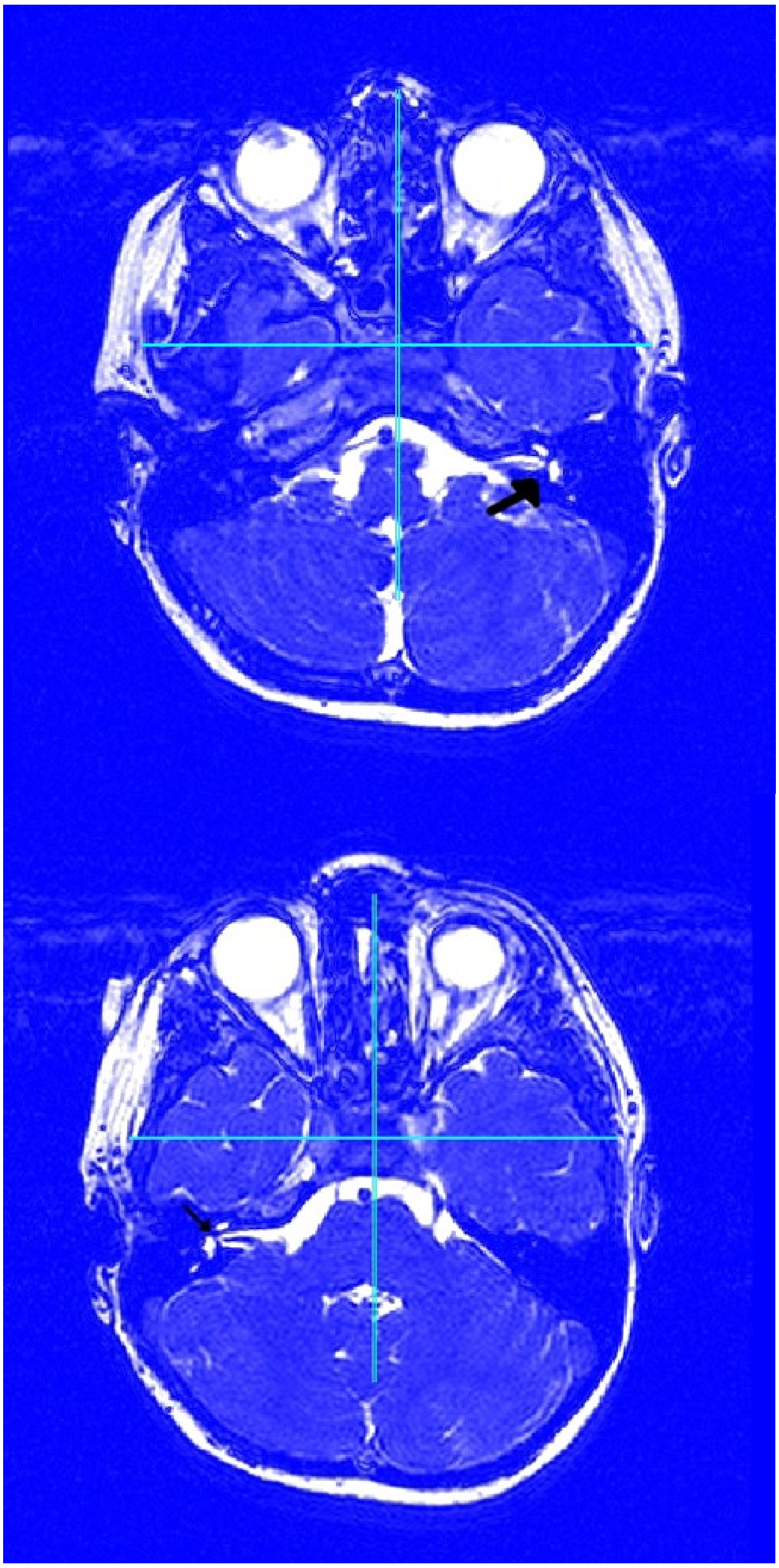

Figure 3A & 3B : Selection of P& P' points

The junction between the auditory meatus and the three vestibular semi-circular canals are selected as markers to measure the spatial orientation of the two PBA petrous parts. The black arrows indicate these junctions, named P (right) (3B) & P' (left) (3A) situated inside the PBA. They are precise and easy to find. The asymmetries of the cerebellum and PBA torque deformation appear on these two axial views.

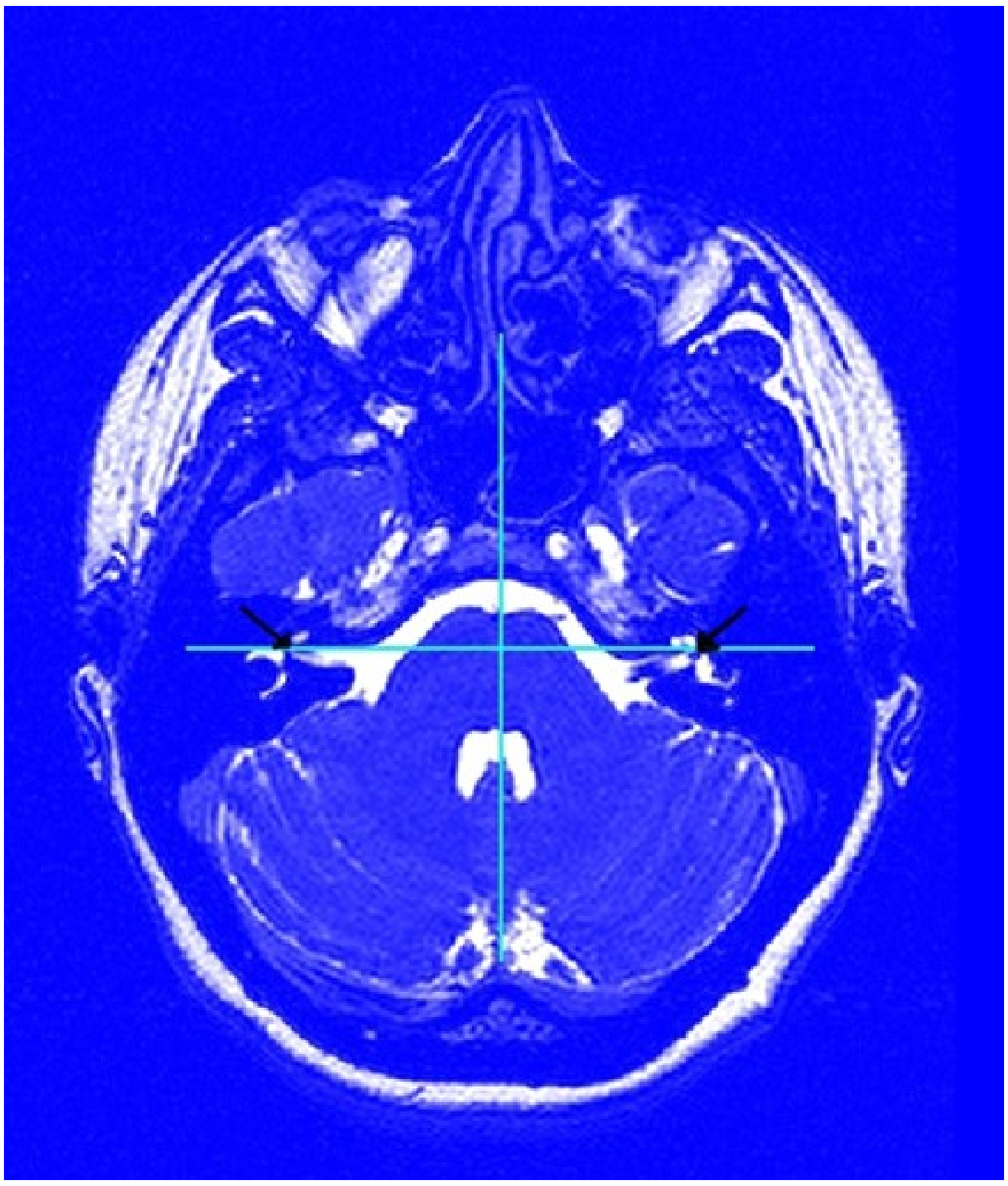

Figure 4 : Symmetrical PBA of a non-IS subject
This control subject presents a symmetrical PBA. The cerebellum is symmetrical. P&P' are located on the same horizontal line.

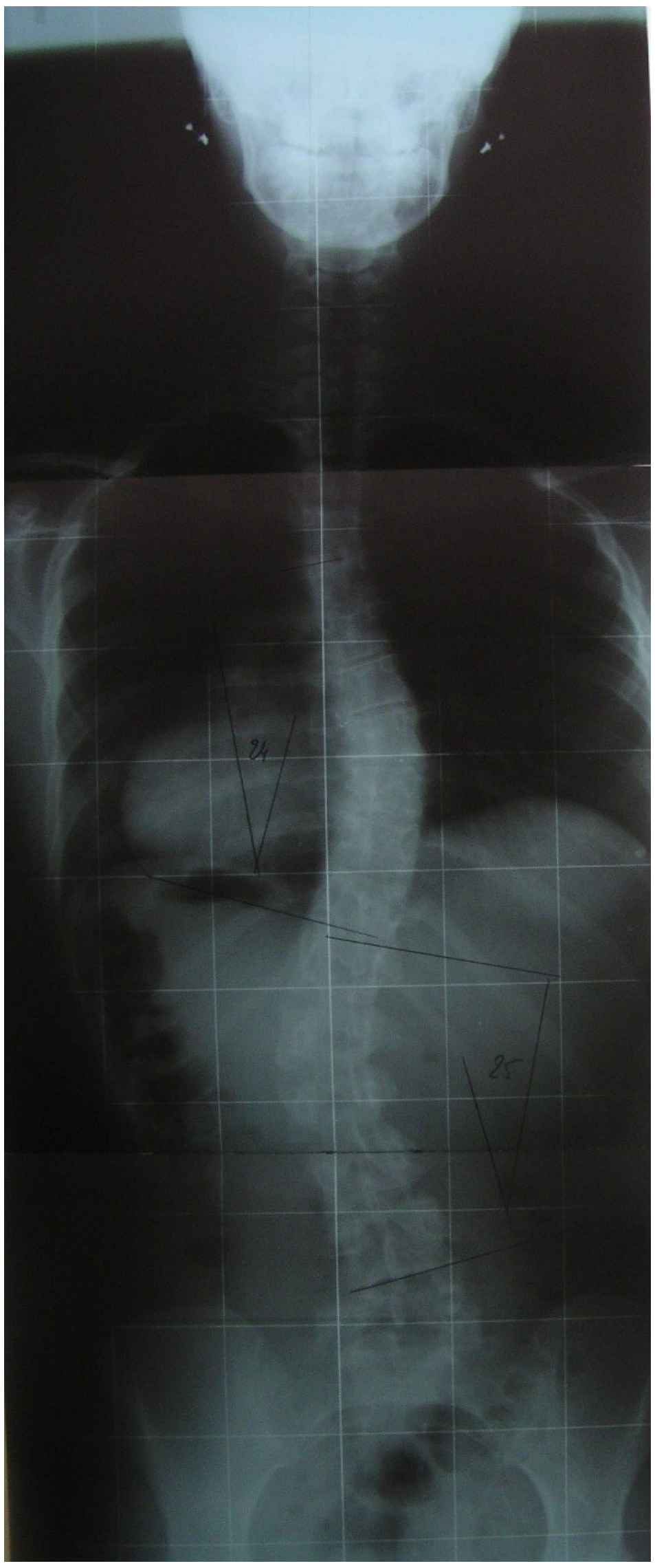

Figure 5 : X-Rays Thoraco-lumbar scoliosis

This patient was the one afflicted with PBA asymmetry detailed in figure 3A & 3B. The asymmetry was calculated with our programme: dP-dP'= - 4. 32mm in relation to the frontal plane, meant that P was backward (AP asymmetry); dP- dP'= - 2.54mm in relation to the NSM plane, meant that P was closer than P' and dP- dP'= - 3,89mmin relation to the horizontal plane, meant that the P was lower than P'. These different results measure the torque deformation of PBA.

# Pba asymmetry and idiopathic scoliosis

Table 1: Statistical comparison between normal and scoliosis Basicranium asymmetries

Gr.: SG: Scoliotic group, CG: Control group,
N: number of subjects
Reference planes: 3 planes of the frame of reference.
Mean L-R: mean value of the difference between P&P' distances (mm) to each of 3 planes (P'=left marker) (P =right marker).
Statistical analysis confirmed that Scoliosis patients presented a significant asymmetry of the posterior basicranium compared to normal subjects especially in AP direction. Non scoliosis

| Gr. | N | Reference planes | Mean L-R | Std Dev | Minimum | Maximum | Median | Lower quart. | Upper quart. | Pr>\|t\| |
|---|---|---|---|---|---|---|---|---|---|---|
| SG | 95 | Frontal (AP) | 5.662 | 2.922 | 0.820 | 16.640 | 4.970 | 3.650 | 7.920 | <.0001 |
| | | Sagittal (lat) | 3.573 | 1.576 | 0 | 10.320 | 3.690 | 2.320 | 4.720 | <.0001 |
| | | Vertical (vert) | 3.516 | 1.591 | 0 | 7.320 | 3.630 | 2.300 | 4.720 | <.0001 |
| CG. | 33 | Frontal (AP) | 2.140 | 1.821 | 0 | 5.450 | 1.960 | 0.590 | 3.030 | <.0001 |
| | | Sagittal (lat) | 1.945 | 1.655 | 0 | 5.560 | 1.830 | 0.710 | 2.640 | <.0001 |
| | | Vertical(vert) | 1.264 | 0.985 | 0 | 3.000 | 1.110 | 0.320 | 2.010 | <.0001 |

subjects presented a weak asymmetry of the posterior base confirming that symmetry is not the rule in Humans.